# A PROBABILISTIC APPROACH TO REDUCE THE ROUTE ESTABLISHMENT OVERHEAD IN AODV ALGORITHM FOR MANET

Preetha K G[1], A Unnikrishnan[2] and K Poulose Jacob[3]

[1]Department of Information Technology, Rajagiri School of Engineering & Technology, Cochin, India
`preetha_kg@rajagiritech.ac.in`
[2]DRDO, Cochin, India
`unnikrishnan_a@live.com`
[3]Department of Computer Science, Cochin University of Science & Technology, Cochin, India
`kpj@cusat.ac.in`

## ABSTRACT

*Mobile Ad-hoc Networks (MANETS) is a collection of wireless nodes without any infrastructure support. The nodes in MANET can act as either router or source and the control of the network is distributed among nodes. The nodes in MANETS are highly mobile and it maintains dynamic interconnection between those mobile nodes. MANTEs have been considered as isolated stand-alone network. This can turn the dream of networking "at any time and at any where" into reality. The main purpose of this paper is to study the issues in route discovery process in AODV protocol for MANET. Flooding of route request message imposes major concern in route establishment. This paper suggests a new approach to reduce the routing overhead during the route discovery phase. By considering the previous behaviour of the network, the new protocol reduces the unwanted searches during route establishment process.*

## KEYWORDS

*Wireless Network, Mobile Network, MANET, AODV, Flooding*

## 1. INTRODUCTION

Data dissemination is the key application domain targeted by MANET. Reliable routing is important and critical in those applications. A major challenge that lies in MANET communication is the unlimited mobility and more frequent failures due to link breakage. Conventional routing algorithms are therefore insufficient for ad-hoc networks. Several researchers have done the qualitative and quantitative analysis of ad hoc routing protocols by means of different performance metrics [ 5,6,8 ]. Researcher Community has broadly classified the routing protocols into two categories [ 2 ]. These are Proactive (Table Driven) routing protocols and Reactive (On Demand) routing protocols. The proactive protocol is the extension of the protocols in wired network. In this case, a routing table is established in each node contains the information of the routes to every other node in the network and this information is updated periodically. Always the up-to date information is available with each node and it is independent of the requirement. However, the periodic propagation of routing information increases the network overhead. A reactive protocol on the other hand finds the route only on demand. Performance analysis shows that reactive protocol outer perform proactive protocol in terms of throughput, PDR and routing overhead [6, 8 ]. So the research has mainly concentrated on reactive protocols.

There are so many reactive protocols have been developed about the last decade. Ad-hoc On-demand Distance Vector (AODV) protocol [ 2 ] is one of the important on demand routing protocol in MANET. In AODV the route is discovered only when it needs to send some data. This on demand characteristics is the major advantage of AODV. Improvements of AODV have been a major attraction in the research work [12,18 ]. The main purpose of the present study is to exploit the route establishment in AODV. Source node initiates the route discovery process by sending route request control packet. The paper also highlights the significance of network overhead during the route discovery process and then tries to propose an idea for reducing the overhead. The proposed technique ensures the flooding of the minimum number of route request message in the network.

The rest of this paper is organized as follows. Section 2 gives how the route discovery process is done using AODV. Section 3 describes issues of flooding the route request messages in the network. Related works are explained in section 4. A new approach to reduce the flooding overhead is discussed in section 5 and a conclusion is given in section 6.

## 2. AN OUTLINE TO ROUTE ESTABLISHMENT IN AODV

Reactive routing protocol eliminates the routing overhead during the periodic information interchange in the proactive routing protocol. In reactive or on demand protocols the up-to date routing table is not retained. It will discover the route only when it is having the data to send. So this reduces the routing overhead. In reactive protocol, the first step of data dissemination is to find the route. This increases the time delay to establish the route. In order to get the advantages of both table driven and on- demand routing protocols, one can combine the approaches to form a hybrid protocol. Various protocols have been developed under these categories over the years.[14]. While AODV, DSR, TORA are the examples of on demand routing protocols and ZRP offers a hybrid protocol. The present paper studies the AODV algorithm and examines the overhead during the route establishment phase. Lots of researches have been reported for the diminution of the unwanted load in the network [3, 4]. The objective of this paper is the study of AODV considering the multiple connections and the overhead required to establish the connections through reroute discovery. The proposed work aims to improve the AODV with the reduction of route discovery overhead.

The basic principle of AODV [2] revolves around the Distance Vector algorithm. In addition to the distance vector algorithm, AODV uses a sequence number to distinguish the route request packet form the old one. This is the extension of DSDV protocol which is table driven. Alternately, the ADOV algorithm minimizes the network broadcast by creating routes on an on demand basis. AODV is classified as pure on demand routing protocol.

Mainly there are two important phases in the AODV implementation viz. route discovery phase and route maintenance phase [1]. In this paper we limit our interest to route discovery phase. To begin with a source node which wants to establish a route, floods the route request message RREQ in the network. Each RREQ packet is associated with a sequence number in order to avoid the duplication. A packet with the new sequence number is only accepted by a receiving node. Any RREQ packet with the same sequence number when received is discarded. If a node has information about the destination then it gives the reply to the source. Otherwise it will send this packet to its neighbours and thus the same request message floods in the network. If there is no reply from any node after a certain interval of time it again tries to connect the route by sending RREQ packet. When a node receives RREQ packet it records the information about its neighbour so as to establish a reverse path. When RREQ packet reaches the destination, it sends a reply message RREP to the sender through the reverse path. Then the source establishes a forward path to the destination. The propagation of route request packet RREQ stops at the destination. The destination node generates the route reply packet RREP and sends to the source node in the reverse direction of the discovered route. This is accomplished by maintaining a route request table at each node that keeps the information about their neighbours that generate the route request message. When a node receives a route reply message, then it checks its route request table and forward the message to the neighbours from which the route request message was received. After establishing the route, the data can be transmitted between source and destination.

During the data transmission, if any link failures revealed the node facing the breakage informs the incident to its neighbours. This message floods in the network and finally reaches to the source node. In this case route establishment process repeats from the beginning. The link failure can be detected by means of HELLO packet, between two nodes. The failure of receiving HELLO packet indicates that the neighbour has moved away. The nodes in MANET are highly mobile so link breakage and link establishment are very frequent. In order to cope up with this problem, each route is associated with a timer. For establishing a better path after a certain interval of time the existing route is deleted and the route establishment process is repeated.

## 3. OVERHEAD DURING ROUTE DISCOVERY PROCESS

One advantage of AODV over other proactive routing protocols like DSDV is the reduction of overhead due to frequent exchange of routing table. AODV minimize this overhead by means of its on demand characteristics. Whenever a data to be send then the route establishment process is carried out. Otherwise no exchange of information between neighbors. Route discovery process starts at source node and ends at the destination node. In AODV flooding is the most suitable strategy for route discovery because of its ever changing topology. Even if the routing overhead is reduced flooding generates huge number of messages. In route discovery phase source node generate the route request control message RREQ and send this to its neighbors. Each node floods the same message over the network, every time it wants to find the destination. Some of the nodes, though just floods the message, but often do not even get the destination. This poses serious challenge in AODV route discovery. Because this unnecessary and unproductive messages, the valuable and limited resources like node energy, bandwidth and battery power of each node are wasted. The problem is more serious with networks of larger sizes.

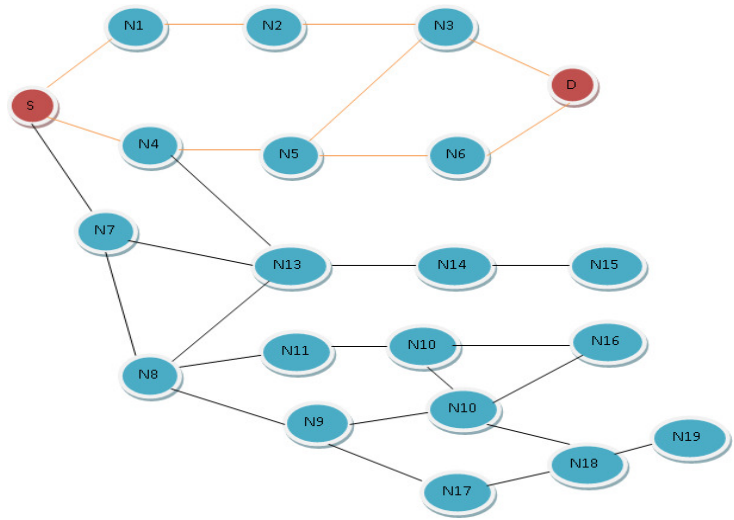

Figure1. A typical network showing the flooding overhead; though there are only three paths from S to D, the flooding is wide spread.

The Fig. 1 shows a typical mobile network with large number of nodes. S denotes source node and D denotes destination node. Initially connection establishment algorithm is started at S by creating RREQ message. S sends this control message to N1, N4 and N7. In initial phase there is no information of destination at each node is available so it again forwards this packet to N2, N13 and N8. But N4 and N7 forward the same packet to N13. One message is redundant for N13. Channel contention is the result of those redundant messages. This also increases the level of congestion and collision in the network.

The route discovery process is initiated when S detect any link breaks. If any node is move away from the network then the neighboring nodes detect this link failure and create a route error message. This control message is forwarded to its neighbors and finally reaches at the source node. When node S receives the route error message then it initiates route discovery process for D. Each time flooding creates extra overhead and the cost of the network become high. An efficient route searching strategy that minimizes the overhead due to excessive flooding is highly desirable. The key design objective of the proposed route searching scheme is to generate the route, with the use of minimum amount of network resources.

## 4. RELATED WORK

Mobile ad hoc network is increasingly promising area of research but they increase the cost of the network in many cases. Lot of research work has been reported on AODV protocol for the reduction in routing overhead [3, 4, 18, 25]. Many of them concentrate on the routing efficiency rather than route discovery.

As was discussed in Sec.3, the flooding in the route discovery phase leads to excessive overhead, with unproductive message flooding through the network. Channel contention, collision, bandwidth overhead, high network cost are the consequences of the excessive flooding. Several schemes are proposed to alleviate the problems due to this unwanted control messages. There exist two common mechanisms used to suppress the propagation of route search overhead. These are query quenching and expanding ring search [1]. In the first method, on receiving RREQ packet the intermediate node will respond, only if it has the information about the destination. So the message flooding can stop at the intermediate node. However, this method does not guarantee the optimal route. In the second method, route discovery attempt is limited some scope. At each attempt, searching scope is increased by some factor. After a maximum threshold of the scope only, the RREQ message is flooded. The drawback of this method is the increase in routing latency.

When there is failure in route discovery or route breaks a new route discovery cycle is repeated. Some of the existing schemes for avoiding duplicate message flooding are following.

1. Probabilistic based [1] – Each node broadcast the message that is received for the first time with some fixed probability P.
2. Counter Based [1] -- Each node broadcast the message only if it has not seen more than c times. If the message has been forwarding several times then extra overhead is induced in the network.
3. Distance based [1] -Each node broadcasts the message only if the physical distance between the nodes is less than some prescribed value d. Distance can be measured by monitoring the signal strength.
4. Location based [1, 4] - A message is broadcast only if the bounding area around the node is less than a. The authors proposed a method to reduce the route discovery overhead by taking into account the location information of the destination node. In this paper the protocols limit the search for a route to the request zone. Request zone is determined based on the expected location of the destination node at the time of route discovery. The size of the request zone is proportional to the average speed of movement, the last location of the destination recorded.

5. Cluster based- Nodes are grouped into clusters; only cluster heads broadcast the control message.
6. Avoiding malicious flooding [3] - One of the basic problem encountered in ad hoc networks is the excessive flooding due to malicious nodes in the network. In the approach reported in [3], the flooding of packets in the network by malicious nodes, who do not obey any limits, is reduced by limiting hyper activism. Each of the receiver nodes has an upper limit of the RREQ it can receive. When a RREQ request reaches a node, it first checks the RREQ_RATELIMIT. If it exceeds the limit, then it avoids all the requests generated from the sending node, which is suspected to be malicious, for the current interval.

## 5. PROPOSED APPROACH TO REDUCE THE ROUTE REQUEST FLOODING

Route request flooding is a major concern in route request phase in AODV. It is possible that the destination is unreachable or request lost due to transmission errors. In these cases, source initiates the route discovery process again. Even though routing overhead is less compared to proactive protocols, route discovery overhead leads to wastage of the limited resources in the network. The main intention of this paper is to propose a simple method, having less overhead and less searching latency for route search in AODV. The proposal also has minimal computational complexity and communication overhead. This method considers the probability of success to connect to the destination. The probability depends on the previous behaviour of a node to get the destination, through an outgoing link. To calculate this probability, connectivity index ($\mu k$) is used as the probability of choosing the neighbour to initiate a route request.

$$\mu_k = \frac{\text{Number of Success Obtained}}{\text{Number of Attempts Made}} = \frac{S_{1\ldots k}}{A_{1\ldots k}}$$

For each attempt, each node updates the $\mu_k$ for each outgoing link using

$$\mu_k \leftarrow \mu_k \, \alpha + (1 - \alpha) \, \mu_{k-1}$$

where **α is a constant, 0< α<1.**

In initial attempt each node's connectivity index on each outgoing link is considered as 1. Second time onwards μk is calculated in each attempt. A threshold value for μk is assigned as .5. Any μk >.5 is considered as eligible to be explored for connectivity. It can be seen that the buildup of μk depends upon α and could be often slow. Accordingly, the first few route requests are made on all the outgoing links. Subsequently, the most favored links alone are chosen corresponding to the higher values of μk.

In figure1 there are many routes from source S to destination D. These routes identified are <S,N1,N2,N3,D>, <S,N4,N5,N6,D>, <S,N4,N5,N3,D>. Initially s creates the route request packet and sends to its neighbors. The Intermediate node has no history of destination node. Then it again sends the packet to its neighbors. This process continues till it reaches the destination. During this route search process each node flood the request packet to its neighbors. Even if some of the node never reaches the destination it also takes part to flood the message. In the figure1 node 17 gets the control message form node S and node13 will get the message from node 4. But that node does not find the destination. So those nodes are unwanted in route searching process. In order to avoid this unwanted searching each node maintains a connectivity index table. Sample connectivity index table of three nodes are given in table1, 2 and 3. In the 11th iteration node 7 does not forward route request message to its neighbors N13 and N8. This way we can block the unwanted routing of route request messages. In case any new link is created after some time then this is informed by the neighboring nodes. If the new link is creating a route to the destination then the intermediate nodes in the path slightly increase their connectivity index of the neighboring nodes.

This approach always considers the previous behaviour of each node to detect the destination node in the route discovery process. This method ensures the maximum reduction of route request message flooding in the network. By updating the connectivity index table each node is aware of the connection potential of the requested destination. With this awareness the node blocks the unwanted forwarding of RREQ control message generated by the source node.

Table 1. Connectivity index table of node S after 10 iterations

| Neighboring Node | $\mu_k$ |
|---|---|
| 1 | 1 |
| 3 | .6 |
| 7 | 0 |

Table 2. Connectivity index table of node 7 after 10 iterations

| Neighboring Node | $\mu_k$ |
|---|---|
| 13 | 0 |
| 8 | 0 |

Table 3. Connectivity index table of node 4 after 10 iterations

| Neighboring Node | $\mu_k$ |
|---|---|
| 5 | .7 |
| 13 | 0 |

## 6. CONCLUSIONS

It has been observed that flooding is the simple mechanism of route searching in an on demand routing algorithm but this wastes network resources, induces congestion, contention and bandwidth overhead in the network. The new approach has been designed to minimize the flooding overhead to a given extent. In the new approach each node forward the message to its neighbours depend on the probability of detecting the destination node. The improvement in AODV algorithm outlined above is always enhancing the performance in connection establishment process. This will have a clear performance enhancement in terms of bandwidth, battery power and node energy. This provides a method of less computational complexity and a better communication capability.

.**REFERENCES**

[1]   George Aggelou, "Mobile Adhoc Networks", Tata McGraw-Hill, 2009 ISBN-13:978-0-07-067748-7.
[2]   Perkins CE, Royer EM "Ad-hoc on-demand distance vector routing". In the Proceedings of IEEE WMCSA, pp. 90 –100, 19993.
[3]   Jayesh Kataria, P.S. Dhekne, Sugata Sanyal,"Ad Hoc On-Demand Distance Vector Routing with Controlled Route Requests", International Journal of Computers, Information Technology and Engineering (IJCITAE), Vol. 1, No. 1, pp 9-15, June 2007, Serial Publications
[4]   Young-Bae Ko and Nitin H. Vaidya," Location-Aided Routing (LAR) in mobile ad hoc networks", Wireless Networks 6 (2000) 307–321, Volume 14, Number 3

[5]     J. Broch, D.A. Maltz, D.B. Johnson, Y.-C. Hu, and J. Jetcheva, "A Performance Comparison of Multi-Hop Wireless Ad Hoc Network Routing Protocols," Proc. MobiCom, pp. 85-97, 1998.

[6]     Hong Jiang, Garcia-Luna-Aceves, J.J. "Performance comparison of three routing protocols for ad hoc networks". Proceedings of the tenth International Conference on Computer communications and Networks, Oct 2001, pp. 547-554

[7]     Charles E. Perkins, Elizabeth M. Royer, "Ad-Hoc on Demand Distance Vector Routing." Prac. 2nd IEEE FVksp. Mobile Computing and Applications, pp. 90-100, Feb. 1999

[8]     Matulya Bansal and Gautam Barua, "Performance Comparison of Two On Demand Routing Protocols for Mobile Ad hoc Networks" IEEE Personal Communication, 2002

[9]     G.S. Tomar; Member IEEE, "Modified Routing Algorithm for AODV in Constrained conditions", Second Asia International Conference on Modelling & Simulation, IEEE, 2008

[10]    Kumar Viswanath, "The Adaptive Routing For Group Communications In Multi-Hop Ad-Hoc Networks", The Dissertation June 2005, University Of California, Santa Cruz.

[11]    Liansheng Tan; Peng Yang; Chan,S ,"An Error-Aware And Energy Efficient Routing Protocol In Manets". Computer Communications and Networks, 2007, ICCCN 2007.

[12]    Ting Liu Kai Liu," An Improved Routing Protocol in Mobile Ad Hoc Networks", IEEE 2007 International Symposium on Microwave, Antenna, Propagation, and EMC Technologies For Wireless Communications

[13]    Hubaux, J.-P. Gross, T. Le Boudec, J.-Y. Vetterli, M.: Toward self-organized mobile ad-hoc networks: the terminodes project, Communications Magazine, IEEE, 2001

[14]    Royer, E.M. Chai-Keong Toh: A review of current routing protocols for ad-hoc mobile wireless networks, Personal Communications, IEEE, 1999

[15]    Mano Yadav, Vinay Rishiwal and K. V. Arya," Routing in Wireless Adhoc Networks: A New Horizon", Journal of Computing, Volume 1, Issue 1, December 2009,   ISSN: 2151-9617,HTTPS://SITES.GOOGLE.COM/SITE/JOURNALOFCOMPUTING

[16]    C. Perkins, et al , "Ad Hoc On Demand Distance Vector (AODV) Routing", draft-ietf-manet-aodv-10.txt, Jan 19, 2002.

[17]    C. K. Toh, " A Novel Distributed Routing Protocol to Support AdHoc Mobile Computing", Proceedings of the 1996 IEEE Fifteenth Annual International Phoenix Conference on computers and Communications, pp. 480   486, March 1996

[18]    Espes, D.; Teyssie, C., Approach for Reducing Control Packets in AODV-Based MANETs, Universal Multiservice Networks, 2007. ECUMN '07. Fourth European Conference on Feb. 2007 Page(s):93 - 104

[19]    Frikha, Mounir; Ghandour, Fatma," Implementation and Performance Evaluation of an Energy Constraint Routing Protocol for Mobile Ad - Hoc Networks Telecommunications", AICT 2007. The Third Advanced International Conference on May 2007

[20]    Rehman, Habib-ur, Wolf, Lars, "Performance Enhancement in AODV with Accessibility Prediction Mobile Ad hoc and Sensor Systems", MASS 2007. IEEE International Conference on 8-11 Oct. 2007

[21]    Safa, Haidar; Artail, Hassan; Karam, Marcel; Ollaic, Hala; Abdallah, Rasha, "HAODV: a New Routing Protocol to Support Interoperability in Heterogeneous MANET Computer Systems and Applications", AICCSA '07., IEEE/ACS International Conference on 13-16 May 2007

[22]    Papadimitratos P, Haas J, Sirer EG, Path set selection in mobile ad hoc network, Proceedings of IEEE MobiHoc, pp. 1–11, 2002

[23]    Youngrag Kim, Shuhrat Dehkanov, Heejoo Park, Jaeil Kim, Chonggun Kim, "The Number of Necessary Nodes for Ad Hoc Network Areas ", 2007 IEEE Asia-Pacific Services computing Conference

[24]    C.Siva Ram Murthy and B.S.Manoj, "Ad hoc Wireless Networks", Pearson 2005.ISBN 81-297-0945- 7

[25]    Srdjan Krco and Marina Dupcinov, "Improved Neighbour Detection Algorithm for AODV Routing Protocol", IEEE COMMUNICATIONS LETTERS, VOL. 7, NO. 12, DECEMBER, 2003

International Journal of Distributed and Parallel Systems (IJDPS) Vol.3, No.2, March 2012

**Authors**

Ms. Preetha K G

Ms. Preetha K G has completed her B Tech and M Tech Degree in Computer Science from Calicut University and Dr. MGR University respectively. She is associated with Rajagiri School of Engineering & Technology as an Assistant Professor in the department of Information Technology. She has around twelve years of academic experience. Currently she is a research scholar in Cochin University of science and Technology. Her Research interests includes Mobile Computing, WirelessNetworks, Ad-hoc Networks Etc.

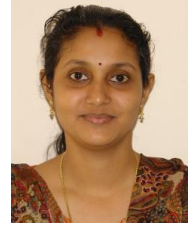

Dr. A Unnikrishnan

Dr. A Unnikrishnan- Graduated from REC (Calicut), India in Electrical Engineering(1975), completed his M.Tech from IIT,Kanpur in Electrical Engineering (1978) and Ph.D from IISc, Bangalore in "Image Data Structures" (1988). Presently, he is the Associate Director in Naval Physical and Oceanographical Laboratory, Kochi which is a premiere laboratory of defence research and development organisation. His field of interests includes Sonar Signal Processing, Image Processing and Soft Computing. He has authored about fifty National and International Journal and Conference Papers. He is a Fellow of IETE & IEI, India.

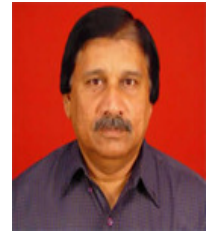

Dr, K Paulose Jacob

Dr. K.Poulose Jacob, Professor of Computer Science at Cochin University of Science and Technology since 1994, is currently Director of the School of Computer Science Studies. . He was Chairman of two Boards of Studies earlier. He has been the Dean of the Faculty of Engineering and is presently Chairman, Board of Studies in Computer Science. He is a member of Academic Council and has been in the University Senate for 10 years. He has presented research papers in several International Conferences in Europe, USA, UK, Australia and other countries. He has delivered invited talks at several national and international events.

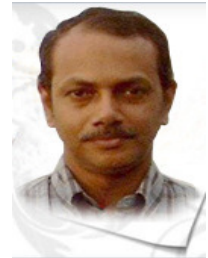

He has served as a Member of the Standing Committee of the UGC on Computer Education & Development. He is the Zonal Coordinator of the DOEACC Society under the Ministry of Information Technology, Government of India. He serves as a member of the AICTE expert panel for accreditation and approval. Dr. K.Poulose Jacob is a Professional member of the ACM (Association for Computing Machinery) and a Life Member of the Computer Society of India.